\documentclass[aps, pra, superscriptaddress, twocolumn, preprintnumbers, amsmath, amssymb, notitlepage, nobalancelastpage, 10pt, longbibliography, colorlinks=true, citecolor=blue, linkcolor=blue, urlcolor=blue]{revtex4-2}

\usepackage{orcidlink}
\usepackage{amsmath}
\usepackage{verbatim}
\usepackage{graphicx}
\usepackage{color}
\usepackage{physics}
\usepackage{yfonts}
\usepackage{soul}
\usepackage[normalem]{ulem}
\usepackage{dsfont}

\newcommand{\cop}{\hat{c}}
\newcommand{\cdop}{\hat{c}^\dag}
\newcommand{\Hop}{\hat{H}}

\newcommand{\pa}[1]{\left( #1 \right)}
\newcommand{\co}[1]{\left[ #1 \right]}

\usepackage{hyperref} 
\usepackage{subfigure}
\usepackage{color}
\definecolor{darkgreen}{rgb}{0,0.60,.2}

\begin{document}

\title{Probing spontaneously symmetry-broken phases with spin-charge separation through noise correlation
measurements}

 \author{Kerman Gallego-Lizarribar,\orcidlink{0009-0002-1853-4660}}
\affiliation{Departament de Física, Universitat Politècnica de Catalunya, Campus Nord B4-B5, 08034 Barcelona, Spain}
\author{Sergi Juli\`a-Farr\'e\,\orcidlink{0000-0003-4034-5786}}
\affiliation{ICFO - Institut de Ciencies Fotoniques, The Barcelona Institute of Science and Technology, Av. Carl Friedrich Gauss 3, 08860 Castelldefels (Barcelona), Spain}
\author{Maciej Lewenstein\,\orcidlink{0000-0002-0210-7800}}
\affiliation{ICFO - Institut de Ciencies Fotoniques, The Barcelona Institute of Science and Technology, Av. Carl Friedrich Gauss 3, 08860 Castelldefels (Barcelona), Spain}
\affiliation{ICREA, Pg. Llu\'is Companys 23, 08010 Barcelona, Spain}
\author{Christof Weitenberg\,\orcidlink{0000-0001-9301-2067}}
\affiliation{Department of Physics, TU Dortmund University, 44227 Dortmund, Germany}
\affiliation{Institut f\"ur Quantenphysik, Universit\"at Hamburg, 22761 Hamburg, Germany}
\affiliation{The Hamburg Centre for Ultrafast Imaging, 22761 Hamburg, Germany}
\author{Luca Barbiero\,\orcidlink{0000-0001-9023-5257}}
\affiliation{Institute for Condensed Matter Physics and Complex Systems,
DISAT, Politecnico di Torino, I-10129 Torino, Italy}
\author{Javier Arg\"uello-Luengo\,\orcidlink{0000-0001-5627-8907}}
\email{javier.arguello.luengo@upc.edu}
\affiliation{Departament de Física, Universitat Politècnica de Catalunya, Campus Nord B4-B5, 08034 Barcelona, Spain}
\affiliation{ICFO - Institut de Ciencies Fotoniques, The Barcelona Institute of Science and Technology, Av. Carl Friedrich Gauss 3, 08860 Castelldefels (Barcelona), Spain}

\begin{abstract}
Spontaneously symmetry-broken (SSB) phases are locally ordered states of matter characterizing a large variety of physical systems. Because of their specific ordering, their presence is usually witnessed by means of local order parameters. Here, we propose an alternative approach based on statistical correlations of noise after the ballistic expansion of an atomic cloud. We indeed demonstrate that probing such noise correlators allows one to discriminate among different SSB phases characterized by spin-charge separation. As a particular example, we test our prediction on a 1D extended Fermi-Hubbard model, where the competition between local and nonlocal couplings gives rise to three different SSB phases: a charge density wave, a bond-ordering wave, and an antiferromagnet. Our numerical analysis shows that this approach can accurately capture the presence of these different SSB phases, thus representing an alternative and powerful strategy to characterize strongly interacting quantum matter.
\end{abstract}

\maketitle
\emph{Introduction.}--
Symmetries play a central role in the characterization of the microscopic properties of the large majority of quantum systems~\cite{Ludwig1996,Beekman2019,FROHLICH2024158}. In this regard, the Mermin-Wagner-Hohenberg theorem~\cite{Hohenberg1967,Mermin1966} demonstrates that, under specific conditions, interacting processes can lead to the formation of the, so called, spontaneously symmetry-broken (SSB) phases. Here, the mechanism of symmetry breaking manifests in the appearance of locally ordered  states of matter that are captured by specific local order parameters (LOPs)~\cite{Landau_ssb}. While theoretical analysis made an extensive use of LOPs, the experimental characterization of SSB phases represented a more challenging task. Nevertheless, the advent of ultracold atomic quantum simulators~\cite{lewenstein2017,Bloch2008} allowed for the investigation of SSB regimes to finally flourish, as proved by the detection of equilibrium~\cite{suDipolar2023} and out-of-equilibrium~\cite{Zahn2022} density waves, supersolids~\cite{tanzi2019,guo2019,chomaz2019} and antiferromagnets~\cite{Mazurenko17,scholl_quantum_2021,shao2024}. In this regard, two main aspects of ultracold experimental platforms proofed particularly important: the impressive versatility in the engineering of Hamiltonians, and the highly accurate detection techniques that allows one to probe local ordering. Specific to this last point, quantum gas microscopy~\cite{Gross2021} represents an extremely complex and powerful method capable of probing both density~\cite{Bakr2009,Sherson2010} and spin~\cite{Boll2016,Mazurenko17} local distributions, where first fundamental results~\cite{hilkerRevealing2017,Vijayan2020} have been obtained. Regarding the tailored interactions, the study of SSB states would benefit from sizeable non-local interactions, whose engineering with dipolar atoms requires the use of lattices with ultrashort lattice spacing~\cite{baierExtended2016,su2023}. Although interesting proposals are present~\cite{fraxanetTopological2022,Sohmen2023}, the diffraction limit might drastically challenge the effectiveness of quantum gas microscopes to detect these phases, which leads to an open quest for less demanding detection schemes. 

\begin{figure}
    \centering    \includegraphics[width=1\columnwidth]{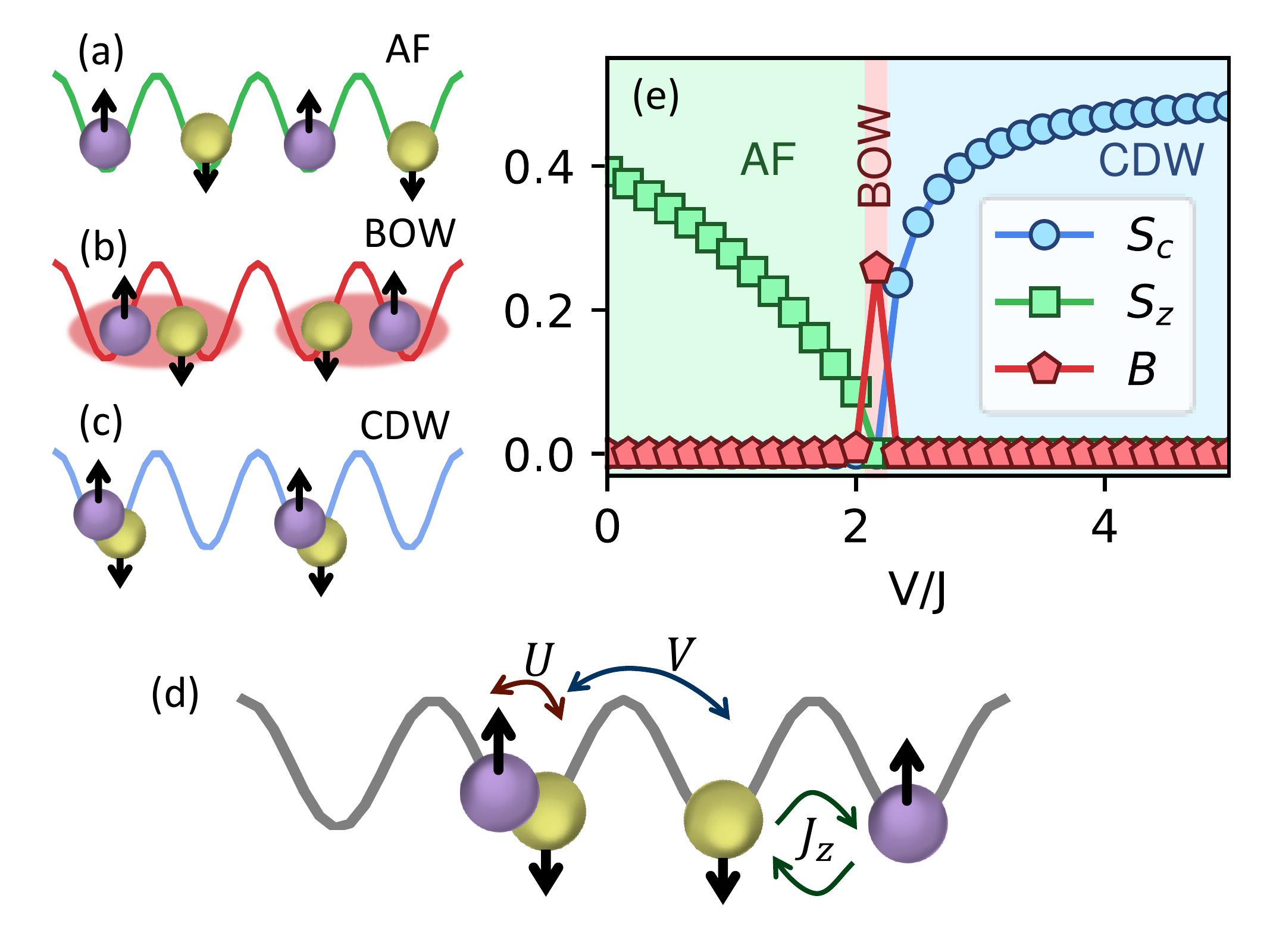}
    \caption{Illustration of the AF (a), BOW (b), CDW (c) phases. (d) Schematic representation of the 1D EFH model~\eqref{eq:extendedHubbard} at half occupation. Atoms of different spin are illustrated with different colors and two-point arrows indicate on-site ($U$) and nearest-neighbor ($V$) interactions. An additional term ($J_z$) couples the spin of neighbor sites. (e) Phase diagram associated to the ground-state of the EFH in the thermodynamic limit. As the value of $V/J$ increases, one observes the transition between the AF, BOW, and CDW phases (coloured in green, red and blue, respectively). See Ref.~\cite{SM} for further details. \emph{Parameters:} $U=4J$ and $J_z=0.5J$.}
    \label{fig:phases}
\end{figure}

Techniques based on time-of-flight measurements offer an alternative, as they allow the extraction of noise correlation measurements (NCMs) from spatial density-density correlations after a ballistic expansion of the gas~\cite{altmanProbing2004}, which does not require a single-site imaging of the lattice. Through such technique, different states of matter have already been efficiently detected~\cite{follingSpatial2005,Spielman2007,Carcy2019,romFree2006,messerExploring2015}. 
In this Letter, we demonstrate that NCMs can as well result highly effective to reveal the presence of SSB phases that are characterized by spin-charge separation~\cite{Tomonaga1950,Luttinger1963,Giamarchi2004}. Specifically, we first derive the expressions that show that NCMs are able to capture the three possible SSB phases occurring in 1D spinful fermionic systems, which are always characterized by gapped charge and spin excitation spectra~\cite{nakamuraTricritical2000,barbieroHow2013}. These are a charge density wave (CDW), an antiferromagnet (AF) and bond-order-wave (BOW) with broken site inversion symmetry, as represented in Fig.~\ref{fig:phases}(a-c). Notably, while the local ordering of CDW and AF appears at the level of lattice sites, in the BOW it takes place in bonds connecting consecutive sites, whose detection is the subject of ongoing efforts for quantum gas microscopy~\cite{impertroLocal2024, nascimbeneExperimental2012}. The alternative use of NCMs to detect this phase had not been proposed before.
Based on such a fundamental aspect, we then test our predictions on an extended Fermi-Hubbard model (EFH) where the aforementioned regimes can be engineered. Here, our numerical analysis for system sizes similar to those of current experiments~\cite{suDipolar2023} demonstrates that NCMs provide a ground state characterization that accurately agrees with the one derived through LOPs, while not relying on spatially resolving the optical lattice. Our results thus provide new insights towards a more complete understanding and characterization of SSB phases of matter.

\emph{NCM for SSB.--} The NCM can be accessed by a sudden release of the optical trap and a posterior fluorescence measurement once the atoms, of mass $m$, have expanded beyond the characteristic size of the lattice during a finite time $\tau$~\cite{altmanProbing2004}. After this ballistic expansion, each momentum is associated to detection in position $x_\nu=\hbar p_\nu \tau/m$, where $p_\nu= p+2\nu k $, $\nu$ is an integer number and $k=2\pi/\lambda$ depends on the wavelength of the lattice geometry. The Pauli principle then prevents the simultaneous detection $\ev{\hat n_\sigma(x) \cdot \hat n_\sigma(x') }$ at distances, $d=x'-x$, which are multiples of $\ell=(2\hbar k)\tau/m$. After normalizing by the case of independent detection $\ev{\hat n(x) }\ev{\hat n(x') }$, the NCM writes as
 \begin{align}
\label{eq:Cd_after}
    \mathcal{N}(d)=1-\frac{\int dx \ev{\hat n(x+d/2) \cdot \hat n(x-d/2) }}{\int dx \ev{\hat n(x+d/2)}\ev{ \hat n(x-d/2) }}\,,
\end{align}
where the bracket notation, $\ev{\cdot}$, indicates the statistical averaging over the region where fluorescence is detected. In this work, whenever we omit explicitly the spin index we refer to the sum over both spins, $\hat n=\hat n_\uparrow + \hat n_\downarrow$.

Interestingly, the presence of peaks in the NCM can reveal symmetries of the state that are associated to structural order in the chain~\cite{guarreraNoise2008,bachCorrelations2004}, thus unveiling the presence of SSB phases. One of these examples is the CDW, which is characterized by a broken translational symmetry that manifests as a perfect alternation between empty and doubly occupied sites. Notably, charge excitations become gapped due to density modulation, while the onsite pairing also generates gapped spin excitations. In a bipartite lattice picture, this manifests as a different site occupation, taking values $n_{e/o,\sigma}$ on each even/odd site for spin $\sigma \in \{\uparrow,\downarrow\}$, which results in a NCM of the form~\cite{SM}
 \begin{equation}
 \mathcal{N}(\nu\ell/2)=\sum_{\sigma}[n_{e\sigma}+(-1)^\nu n_{o\sigma}]^2/[\sum_\sigma(n_{e\sigma}+n_{o\sigma})]^2.
 \label{ncm_densOrd}
 \end{equation}
In analogy to measurements in 2D systems with imposed broken symmetry ~\cite{messerExploring2015}, $\mathcal{N}(\ell/2)$ is null for a homogeneous distribution, and nonvanishing for a bipartite occupation. Therefore, the NCM in Eq. (\ref{ncm_densOrd}) serves as a rigorous probe that is capable of detecting translational symmetry broken phases. In the case of a spinful fermionic CDW occurring in half-filled 1D lattices, we then expect $n_{e\sigma}=1$ and $n_{o\sigma}=0$, so that the NCM saturates to $\mathcal{N}(\ell/2)=0.5$.

 Interestingly, such analysis is also relevant for the characterization of the AF SSB represented in Fig.~\ref{fig:phases}(a). This phase exhibits a finite charge gap that originates from energetically-prevented local pairing, while the perfect spatial alternation between $\uparrow$ and $\downarrow$ particles translates into gapped spin excitations. For the half-filled fermionic system described above, the AF state, $n_{o\uparrow}=n_{e\downarrow}=1$ and $n_{o\downarrow}=n_{e\uparrow}=0$, saturates again the NCM in Eq.~\eqref{ncm_densOrd} to the value $\mathcal{N}(\ell/2)=0.5$.
 
In these two SSB phases, local order is present at the level of lattice sites. In contrast to this, the BOW phase occurring in different fermionic chains \cite{Fabrizio1999,nakamuraTricritical2000,Aligia2007}, is characterized by the ordering of local bonds, $b_{r\sigma}=\ev{\hat c^{\dagger}_{r\sigma} \hat c_{r+1\sigma}+\text{h.c.}}$, which results in a spontaneously generated lattice dimerization [see Fig.~\ref{fig:phases}(b)]. In analogy to AF, the charge gap in the BOW reflects in a uniform distribution of singly occupied lattice sites, while the lattice dimerization gives rise to the formation of singlets in neighboring sites causing spin gapped excitations. Interestingly, the local ordering in the bonds translates into additional terms in the NCM~\cite{SM},

\begin{equation}
\label{eq:Fock_bipartite2}
\mathcal{N}(\nu\ell/2)= \frac{\sum_\sigma \co{\sum_{r}(-1)^{\nu r}n_{r\sigma}}^2+0.5\sum_\sigma  \co{\sum_{r}(-1)^{\nu r}b_{r\sigma}}^2 }{(\sum_{\sigma r}n_{r\sigma})^2-0.5(\sum_{\sigma r}b_{r\sigma})^2}\,.
\end{equation}
In particular, in the SSB BOW we expect a different occupation of even/odd bonds, $n_{e/o,\sigma}=0.5$, $b_{e\sigma}=1$, $b_{o\sigma}=0$, and one obtains a nonvanishing $\mathcal{N}(\ell/2)=0.5$.
Therefore, a non-zero value of $\mathcal{N}(\ell/2)$ in a fermionic chain indicates the presence of any of three possible symmetry breakings depicted in Fig.~\ref{fig:phases}. In the following, we apply these results to a minimal model where the three SSB phases appear, and illustrate a strategy to rigorously detect individually each SSB phase by using NCMs.

\begin{figure*}
    \centering    \includegraphics[width=2.1\columnwidth]{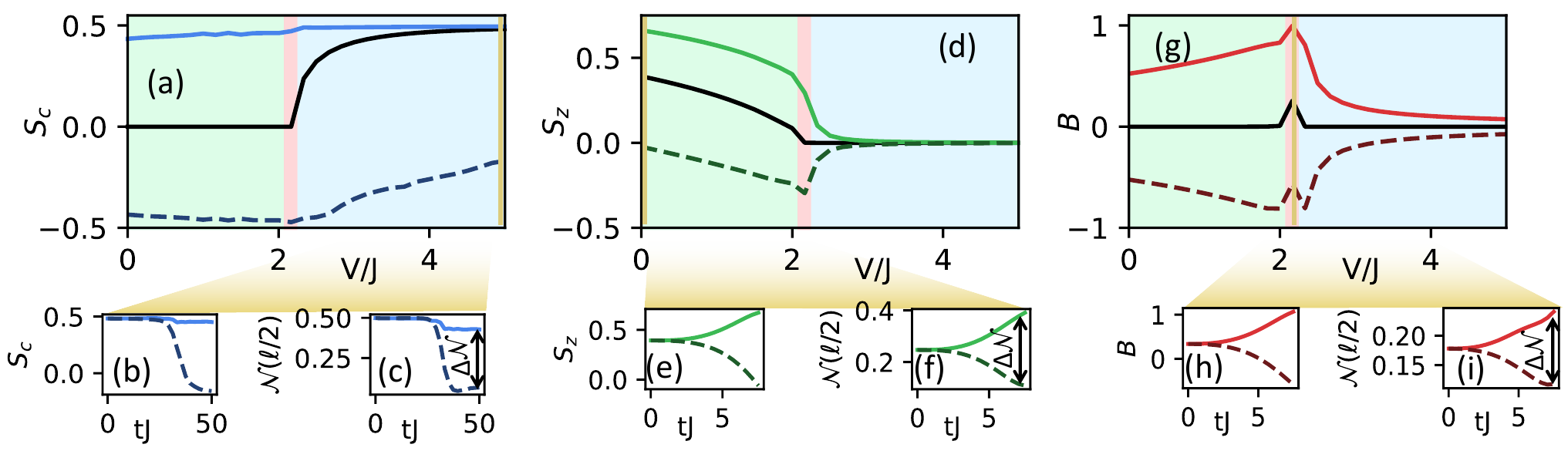}
    \caption{(a) Black line follows the LOP associated to charge order ($S_c$) for different values of nearest-neighbor interaction strength, $V/J$. Blue lines indicate indicate the final value of $S_c$ after one adiabatically introduces the on-site dimerized energy shift in Eq.~\eqref{eq:pathCharge}. Blue continuous line corresponds to the symmetry sector that is compatible with the initial sector of the ground-state, and the dashed line corresponds to the opposite sector. Panels (b) and (c) indicate the evolution of $S_c$ and $\mathcal{N}(\ell/2)$ (respectively) for a fixed $V/J=5$ (yellow line). In (d) we focus on the LOP associated to antiferromagnetic order ($S_z$), represented in black. Continuous and dashed green lines show the final value of $S_z$ for the even and odd adiabatic paths in Eq.~\eqref{eq:pathAnti}. Panels (e) and (f) indicate $S_z$ and $\mathcal{N}(\ell/2)$ (respectively) for $V/J=0$ (yellow line). In (g) we follow the same approach for the LOP associated to bond order ($B$), represented in black. Continuous and dashed red lines show the final value of $B$ after an adiabatic frustration of the tunneling in the bonds corresponding to the transformation~\eqref{eq:pathBond} compatible with the ground-state, or the opposite one, respectively. Panels (h) and (i) indicate the evolution of $B$ and $\mathcal{N}(\ell/2)$ (respectively) for $V/J=2.16$ (yellow line). \emph{Parameters:} $U=4J$, $J_z=0.5J$, $\Delta=10J$, 100 sites, and $TJ=50$.}
    \label{fig:evolution}
\end{figure*}

\emph{SSB in the 1D EFH model.}--
We consider a 1D EFH model describing a chain of length $L$ where $N$ spinful fermions, labeled by $\sigma=\uparrow,\downarrow$, interact through local $U$ and nearest-neighbor $V$ interactions~\cite{baierExtended2016}, and are subject to an antiferromagnetic coupling $J_z>0$ [see Fig.~\ref{fig:phases}(d)]~\cite{baldelliDeconfined2024}: 
\begin{equation}
 \label{eq:extendedHubbard}
\begin{split}
\Hop=&-J\sum_{\langle ij\rangle, \sigma}(\cdop_{i\sigma}\cop_{j\sigma}+\textrm{H.c.}) +V \sum_{\langle ij\rangle}\hat{n}_i\hat{n}_{j}\\
& +U\sum_{i=0}^{L-1}\hat{n}_{i\uparrow}\hat{n}_{i\downarrow} + J_z \sum_{\langle ij\rangle} \hat{S}^z_{i} \hat{S}^z_{j},\\
\end{split}
\end{equation}
where $J$ parametrizes the nearest neighbor hopping, and $\hat{S}^z_i = (\hat{n}_{i\uparrow}-\hat{n}_{i\downarrow})/2$. Here, we fix both the system density $N/L=1$ and the total magnetization $\sum_i \hat{S}^z_i=0$. For weak $V$, the on-site interaction dominates and the finite value of $J_z$ turns out to be responsible for the breaking of the $SU(2)$ spin rotational symmetry, giving rise to the appearance of an AF phase~\cite{Japaridze2000}. A strong non-local repulsion $V$ causes that fermions in neighboring sites become energetically unfavorable, which results in the formation of a CDW with broken  translational symmetry~\cite{nakamuraTricritical2000,Sandvik2004}. For low enough $J_z$, the effective frustration generated in the regime $U\approx2V$ turns out to be responsible for an effective Peierls instability that results in the formation of a SSB BOW~\cite{nakamuraTricritical2000}.

The presence of the aforementioned SSB phases can be characterized by their respective LOP,
\begin{align}
\label{eq:localCharge}
    S_c &=\frac{1}{L}\sum_{i} (-1)^i \ev{\hat n_{i\uparrow}+\hat n_{i\downarrow}}/2\,, \\
\label{eq:localAntiferro}
    S_z &=\frac{2}{L}\sum_{i} (-1)^i \ev{\hat{S}^z_i} \,,\\
\label{eq:localBond}
    B &=\frac{2}{L}\sum_{i\sigma} (-1)^i b_{i\sigma} \,,
\end{align}
capturing the CDW, AF and BOW phases, respectively.
For fixed values $U=4J$ and $J_z=0.5J$, our calculations of the LOPs in Fig.~\ref{fig:phases}(e) confirm that the variation of $V$ result in the appearance of the discussed SSB phases in the ground-state of the EFH model~\eqref{eq:extendedHubbard}~\footnote{For each of those phases, both symmetry sectors are degenerate in the thermodynamic limit. However, in a real experiment one of the symmetry sectors will be preferred by imperfections, the state preparation, and/or finite-size effects. To account for this,the degeneracy breaks in our tensor networks calculation with the choice of the initial state to be optimized (see Ref.~\cite{SM}).}.

\emph{Discriminating among the different SSB through NCMs.}-- 
So far, we have discussed that NCMs based on TOF measurements can probe the three SSB phases that appear in Eq.~\eqref{eq:extendedHubbard}, but cannot directly discriminate among them in the reciprocal space where they operate. To circumvent this situation, we introduce a strategy where NCMs, in combination with tunable superlattices, can be used to reveal the presence of each SSB. For the ground state $\ket{\psi_0}$ of $\Hop$ and a symmetry of interest, we induce a time-dependent superlattice to reduce the energy of either of the possible charge or spin sectors associated to the symmetries of study. We will use the notations $\text{\textcircled{C}}$, $\text{\textcircled{A}}$, and $\text{\textcircled{B}}$ for the superlattice modulation compatible with the order that spontaneously appear in the CDW, AF, and BOW phase, respectively:
\begin{align}
    \label{eq:pathCharge}
    \hat H_{\text{\textcircled{C}}\substack{\text{e} \\ \text{o}}}(t)&=-\Delta_\text{T}(t) \sum_{\sigma,i\in \substack{\text{even} \\ \text{odd}}}  \hat n_{i\sigma}\,,\\
    \label{eq:pathAnti}  \hat H_{\text{\textcircled{A}}\substack{\text{e} \\ \text{o}}}(t)&=-\Delta_\text{T}(t) \sum_{i\in \substack{\text{even} \\ \text{odd}}} \pa{\hat n_{i\uparrow}-n_{i\downarrow}}\,,\\
    \label{eq:pathBond}
    \hat H_{\text{\textcircled{B}}\substack{\text{e} \\ \text{o}}}(t)&= \Delta_\text{T}(t) \sum_\sigma \sum_{i\in \substack{\text{even} \\ \text{odd}}}\co{ \hat c^\dagger_{2i,\sigma}\hat c_{2i+1,\sigma}  + \text{H.c.}} \,.
\end{align}
Here, we have defined $\Delta_{\text{T}}(t)=(t/T)^2\Delta$, where $T$ is the total time of the passage, and the index $e/o$ indicates whether the modulated Hamiltonian favours the occupation of even/odd sites [or bonds in the case of $\text{\textcircled{B}}$].

\begin{figure}[!tbp]
    \centering   
    \includegraphics[width=1\columnwidth]{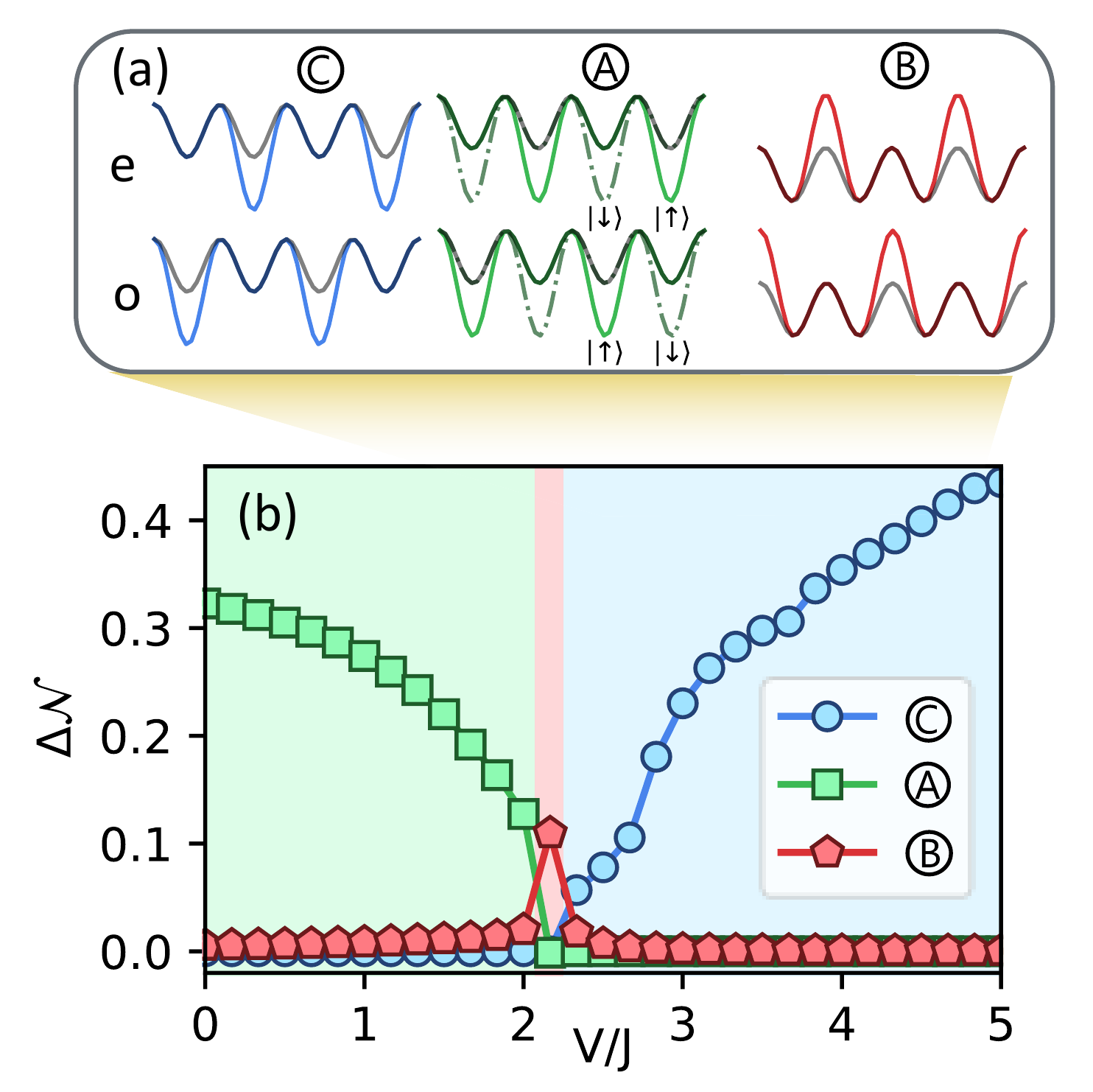}
    \caption{(a) Schematic representation of the superlattices induced by the adiabatic transformations (\ref{eq:pathCharge}-\ref{eq:pathBond}). (b) Difference in the NCM reached after the even or odd sector of those transformations reveal the presence of the CDW, AF and BOW phases of the EFH model~\eqref{eq:extendedHubbard}, respectively (see main text). Parameters as in Fig.~\ref{fig:evolution}.}
    \label{fig:phasesNCM}
\end{figure}

Let us now illustrate in greater detail how this strategy can assist distinguishing the discussed SSB phases. Starting with the CDW phase, 
in Fig.~\ref{fig:evolution}(b,c) we show the evolution of $S_c$ and the noise correlator (respectively), as one starts from the ground-state of the EFH model for a fixed value of $V=5J$, where the CDW phase is present. The initial values $\mathcal{N}(\ell/2)\approx 0.5$ and $S_c\approx 0.5$ indicate the presence of SSB phase where even sites are initially occupied. As we now shape the lattice to decrease the energy-cost of occupying even sites, we observe at final time $TJ=50$ no change in those values along the adiabatic evolution through the ground-state manifold of $\Hop +\hat H_{\text{\textcircled{C},e}}$ (continuous lines), which indicates that the SSB is unaltered. However, along the evolution with $\Hop +\hat H_{\text{\textcircled{C},o}}$ (dashed line) where the occupation of odd sites is favoured, we observe that the state fails to reach adiabatically the state with opposite parity due to the closing of the gap along that path. As a consequence, the final values of the LOP and NCM do not saturate to 0.5, indicating that a state with broken symmetries has not been reached in this occasion. When looking into the final values of $S_c$ after these two different paths in  Fig.~\ref{fig:evolution}(a), we observe that this behaviour holds for $V\gtrsim 2.25J$ (coloured in blue, where the CDW phase is present) while, for $V\lesssim 2.25J$, the evolution along the even (continuous lines) and odd (dashed) adiabatic paths result into the same absolute values of $S_c$. Remarkably, this different evolution along the even and odd paths (\ref{eq:pathCharge}-\ref{eq:pathBond}) schematized in Fig.~\ref{fig:phasesNCM}(a) can also be sensed purely from NCMs. In the blue line of Fig.~\ref{fig:phasesNCM}(b), we calculate the maximum difference $\Delta \mathcal N$ between the values of $\mathcal{N}(\ell/2)$ reached over time by those paths, which is null for $V\lesssim 2.25J$, while it is nonvanishing in the region $V\gtrsim 2.25J$, thus revealing the presence of the CDW phase without any need of LOPs or single-site resolution. 

The applicability of this method is not restricted to CDWs, but it can also be used to detect AF and BOW phases, as we illustrate in Fig.~\ref{fig:evolution}(d-f) and \ref{fig:evolution}(g-i), by repeating the same analysis for the adiabatic passages $  \hat H_{\text{\textcircled{A}}}(t)$, and $  \hat H_{\text{\textcircled{B}}}(t)$, respectively. In the first case, we observe that the unequal evolution of $S_z$ and the NCM [panels (e) and (f)]  under the two parities of the path in Eq.~\eqref{eq:pathAnti} reveal the presence of an AF phase for $V/J \lesssim 2.1$. Starting from the ground-state of the EFH model for the latter case and $V/J= 2.16$, the BOW phases manifests from the different values of the bond-order LOP, and NCMs [panels (h) and (i)] reached by the system after the adiabatic suppression of tunneling in even or odd bonds (indicated with red continuous and dashed lines respectively), following Eq.~\eqref{eq:pathBond}. 

The presence of a specific SSB phase in the Hamiltonian thus manifests as a maximum discrepancy $\Delta \mathcal{N}\neq 0$, between the value of the NCM reached over time by the two possible paths [e) and o)] associated to that symmetry  (\ref{eq:pathCharge}-\ref{eq:pathBond}). One should note that, in an experiment, the SSB will be decided by an uncontrolled pinning potential and will be different in each realization. Therefore, the symmetry breaking will manifest as a bimodal distribution of measurement outcomes, where the separation $\Delta \mathcal{N}$ between the two peaks is the relevant measure. Remarkably, this magnitude, represented in Fig.~\ref{fig:phasesNCM}(b), allows one to reconstruct the same phase diagram as the one obtained from LOP, [see Fig.~\ref{fig:phases}(e)]. 

\emph{Discussion \& outlook.}--
We have shown that noise correlation measurements can represent a fundamental tool in order to probe spontaneously symmetry broken phases with spin-charge separation. Specifically, we derived an alternative detection scheme that combines tailored lattice designs with time-of-flight probings. The latter allowed to accurately reveal the presence of each of the three SSB phases that can occur in 1D fermionic systems. It is worth to underline that proposals aimed to explore the SSB, CDW and BOW phases have been mainly based on trapping magnetic atoms into 1D optical lattices~\cite{DiDio2014,Julia2022}, where the subwavelength separations required to enhance the interaction strength pose a challenge for quantum gas microscopy. In this regard, our proposed scheme, combined with the recently introduced technique of quantum gas magnifier~\cite{Asteria2021}, might thus represent a more feasible and flexible strategy. Other approaches to obtain extended interactions would be the use of Rydberg atoms~\cite{guardado-sanchezQuench2021,weckesserRealization2024} or dipolar molecules~\cite{bigagliObservation2024, rosenbergObservation2022} trapped in optical lattices. Although, these phases do not allow for a description in terms of local order parameters, they are still characterized by the phenomenon of symmetry breaking that our proposed method might detect efficiently. In conclusion, our results open up new avenues in the comprehension and detection of spontaneously symmetry-broken states of matter.

\textbf{Acknowledgments}.
We thank T. Esslinger for useful discussions. The numerical calculations were performed using the TenPy Library~\cite{hauschildEfficient2018}. KGL acknowledges the Catalonia Quantum Academy (CQA). ICFO group acknowledges support from: ERC AdG NOQIA; MICIN/AEI (PGC2018-0910.13039/501100011033, CEX2019-000910-S/10.13039/501100011033, Plan National FIDEUA PID2019-106901GB-I00, FPI; MICIIN with funding from European Union NextGenerationEU (PRTR-C17.I1): QUANTERA MAQS PCI2019-111828-2); MCIN/AEI/ 10.13039/501100011033 and by the “European Union NextGeneration EU/PRTR" QUANTERA DYNAMITE PCI2022-132919 within the QuantERA II Programme that has received funding from the European Union’s Horizon 2020 research and innovation programme under Grant Agreement No 101017733Proyectos de I+D+I “Retos Colaboraci\'on” QUSPIN RTC2019-007196-7); Fundaci\'o Cellex; Fundaci\'o Mir-Puig; Generalitat de Catalunya (European Social Fund FEDER and CERCA program, AGAUR Grant No. 2021 SGR 01452, QuantumCAT \ U16-011424, co-funded by ERDF Operational Program of Catalonia 2014-2020); Barcelona Supercomputing Center MareNostrum (FI-2023-1-0013); EU (PASQuanS2.1, 101113690); EU Horizon 2020 FET-OPEN OPTOlogic (Grant No 899794); EU Horizon Europe Program (Grant Agreement 101080086 — NeQST), National Science Centre, Poland (Symfonia Grant No. 2016/20/W/ST4/00314); ICFO Internal “QuantumGaudi” project; European Union’s Horizon 2020 research and innovation program under the Marie-Sk\l odowska-Curie grant agreement No 101029393 (STREDCH) and No 847648 (“La Caixa” Junior Leaders fellowships ID100010434: LCF/BQ/PI19/11690013, LCF/BQ/PI20/11760031,LCF/BQ/PR20/
11770012, LCF/BQ/PR21/11840013). Views and opinions expressed are, however, those of the author(s) only and do not necessarily reflect those of the European Union, European Commission, European Climate, Infrastructure and Environment Executive Agency (CINEA), nor any other granting authority. Neither the European Union nor any granting authority can be held responsible for them. C. W. acknowleges funding by the Cluster of Excellence “CUI: Advanced Imaging of Matter” of the Deutsche
Forschungsgemeinschaft (DFG) - EXC 2056 - Project ID No. 390715994, by the
DFG Research Unit FOR 2414, Project No. 277974659, and by the European Research Council (ERC) under the European Union’s Horizon 2020 research and innovation program under Grant Agreement No. 802701. L. B. acknowledges financial support within the DiQut Grant No. 2022523NA7 funded by European Union – Next Generation EU, PRIN 2022 program (D.D. 104 - 02/02/2022 Ministero dell’Università e della Ricerca). JAL acknowledges support by the Spanish Ministerio de Ciencia e Innovaci\'on (MCIN/AEI/10.13039/501100011033, Grant No. PID2023-147469NB-C21), and by the Generalitat de Catalunya (Grant No. 2021 SGR 01411).

\clearpage
\onecolumngrid
\appendix
\renewcommand\thefigure{S\arabic{figure}}  
\renewcommand{\theequation}{S\arabic{equation}}
\renewcommand\appendixname{SM}
\setcounter{figure}{0} 
\setcounter{page}{1} 
\section*{Supplemental material}

\section{Noise correlators}
\label{ap:noise_correlation}
The fermionic statistics of the simulating atoms can manifest as an interference in the atomic density distribution when the optical lattice is suddenly removed. Atoms are allowed to expand ballistically over a fixed time, $\tau$, that we consider sufficient to neglect the initial size of the lattice.

After this expansion, measurable correlations in the density distribution,
\begin{align}
\label{eq:Cd_after2}
    \mathcal{N}(d)=1-\frac{\int dx \ev{\hat n(x+d/2) \cdot \hat n(x-d/2) }_\tau}{\int dx \ev{\hat n(x+d/2)}_\tau\ev{ \hat n(x-d/2) }_\tau}\,,
\end{align}
are proportional to the initial momentum distribution before expansion, 
\begin{equation}
    \ev{\hat n_\tau(x)}_\tau  \approx m/(h \tau) \ev{\hat n_{q(x)}}_0\,,
\end{equation}
following the relation, $q(x)=mx/(\hbar \tau)$, where $m$ is the atomic mass. In the following, we will omit the subindex, $_0$, to indicate the initial time before ballistic expansion. At long times, one can then express $\mathcal{N}(d)$ in terms of the initial annihilation(creation) operators $\cop^{(\dagger)}_{r,\sigma}$ of a fermionic atom with internal state, $\sigma$, placed at site, $r$, of the optical lattice before the expansion. 

The density correlations in Eq.~\eqref{eq:Cd_after2} write as~\cite{altmanProbing2004,follingSpatial2005},
\begin{align}
    \ev{ \hat n_\sigma(x) }_\tau  &\approx \frac{1}{W}\sum_{rs}e^{i(r-s)Q(x)}\ev{\cdop_{r,\sigma}\cop_{s,\sigma}}\,,\\
    \label{eq:evnn}
        \ev{ \hat n_\sigma(x) \cdot \hat n_{\sigma'}(x') }_\tau  & \approx  \frac{1}{W^2}\sum_{rsr's'}e^{i(r-s)Q(x)+i(r'-s')Q(x')}\ev{\cdop_{r,\sigma}\cop_{s,\sigma}\cdop_{r',\sigma'}\cop_{s',\sigma'}}\,,
\end{align}
where $Q(x)=q(x)\lambda/2$, and $W=\hbar\tau/(ma_0)$ approximates a Gaussian envelope for the expansion of the initial Wannier functions in one-dimension. 

Exchanging the order of integration in Eq.~\eqref{eq:Cd_after2}, one sees that,
\begin{equation}
\begin{split}
   \int dx \ev{\hat n_\sigma(x+d/2) }\ev{ \hat n_{\sigma'}(x-d/2)}  &\approx \frac{2\pi}{W^2}\sum_{rsr's'} e^{iQ(d)(r-s)}\ev{\cdop_{r,\sigma}\cop_{s,\sigma}}\ev{\cdop_{r',\sigma'}\cop_{s',\sigma'}} \delta[(r-s)+(r'-s')]\,,
   \end{split}
\end{equation}
and for the calculation of $\int dx \ev{\hat n_\sigma(x+d/2) \cdot \hat n_{\sigma'}(x-d/2)}\,,$ we can simplify Eq.~\ref{eq:evnn} as,
\begin{equation}
\label{eq:apC(D)}
\begin{split}
   \sum_{rsr's'} &\ev{\cdop_{r,\sigma}\cop_{s,\sigma}\cdop_{r',\sigma'}\cop_{s',\sigma'}} e^{iQ(d/2)[(r-s)-(r'-s')]}\int dx e^{iQ(x) [(r-s)+(r'-s')]}\\
   &=\sum_{rsr's'} e^{iQ(d)(r-s)}\ev{\cdop_{r,\sigma}\cop_{s,\sigma}\cdop_{r',\sigma'}\cop_{s',\sigma'}} \delta[(r-s)+(r'-s')]]\,.
   \end{split}
\end{equation}

In the case of a Fock state, one can further simplify,
\begin{equation}
\label{eq:fock_approx}
\ev{\cdop_{r,\sigma}\cop_{s,\sigma}\cdop_{r',\sigma'}\cop_{s',\sigma'}}=
\ev{\cdop_{r,\sigma}\cop_{s,\sigma}}\ev{\cdop_{r',\sigma'}\cop_{s',\sigma'}} -\ev{\cdop_{r,\sigma}\cop_{s',\sigma}}\ev{\cdop_{r',\sigma}\cop_{s,\sigma}}\delta_{\sigma,\sigma'}+\ev{\cdop_{r,\sigma}\cop_{s',\sigma}}\delta_{s,r'} 
 \delta_{\sigma,\sigma'}\,,
\end{equation}
and,
\begin{equation}
\label{eq:cd_lattice}
    \mathcal{N}(d)\approx \frac{-\delta(d) L\sum_{r\sigma} \ev{\hat n_{r\sigma}} + \sum_{\sigma}\sum_{rsr's'} e^{iQ(d)(r-s)}\ev{\cdop_{r,\sigma}\cop_{s',\sigma}}\ev{\cdop_{r',\sigma}\cop_{s,\sigma}} \delta[(r-s)+(r'-s')]}{\sum_{\sigma\sigma'}\sum_{rsr's'} e^{iQ(d)(r-s)}\ev{\cdop_{r,\sigma}\cop_{s,\sigma}}\ev{\cdop_{r',\sigma'}\cop_{s',\sigma'}} \delta[(r-s)+(r'-s')]}\,,
\end{equation}
where $L$ is the number of sites.
In order to get some insights, one can look into two particular scenarios.

\begin{itemize}
\item
In the case of a {Fock state}, $\ev{\cdop_{r,\sigma}\cop_{s,\sigma'}}=\delta_{rs}\delta_{\sigma\sigma'}n_{r\sigma},$ and Eq.~\eqref{eq:cd_lattice} simplifies as,
$$    
\mathcal{N}(d)\approx -\frac{\delta(d)}{N/L}+ \frac{1}{N^2}\sum_{\sigma}\sum_{rs} e^{iQ(d)(r-s)}n_{r\sigma}n_{s\sigma}\,,
$$
where $N=\sum_{r\sigma} n_{r\sigma}$ is the total number of atoms.
We can consider the case of a charge density wave with occupation $n_ {e/o,\sigma}$ on even/odd sites. One obtains~\cite{messerExploring2015},    
\begin{equation}
\mathcal{N}(d)\approx -\frac{\delta(d)}{\sum_\sigma 0.5 (n_{e\sigma}+n_{o\sigma} ) }+
\frac{\sum_{\sigma}[n_{e\sigma}^2 + n_{o\sigma}^2+2n_{e\sigma} n_{o\sigma}\cos(Q(d))]}{[\sum_\sigma(n_{e\sigma}+n_{o\sigma})]^2}\,,
\end{equation}
finding a periodic correlation at multiples of the reciprocal lattice distance, $\ell/2=\frac{2\pi}{\lambda}\frac{\hbar \tau}{m}\,,$
\begin{equation}
\label{eq:Fock_bipartite1}
        \mathcal{N}(\nu  \ell /2)\approx 
        \frac{\sum_{\sigma}[n_{e\sigma}+(-1)^\nu n_{o\sigma}]^2}{[\sum_\sigma(n_{e\sigma}+n_{o\sigma})]^2}\,,
\end{equation}
for $\nu \in \mathds{N}$.

A particular example is the {purely antiferromagnetic} state, $|\uparrow\downarrow\uparrow\ldots\rangle$, where, $n_{o\uparrow}=n_{e\downarrow}=1$ and $n_{o\downarrow}=n_{e\uparrow}=0$, so that Eq.~\eqref{eq:Fock_bipartite1} reduces to $\mathcal{N}(\ell/2)=1/2$. Interestingly, this is the same result one obtains for the case of a charge density wave, where $n_{o\sigma}=1$ and $n_{e\sigma}=0$.  Intuitively, as the two spin states do not anticommute with each other and a one-site spatial displacement between the two ordered chains is not resolved in reciprocal space, the antiferromagentic chain is analogous to a purely charge-ordered state.

We should highlight that the simplified form in Eq.~\eqref{eq:Fock_bipartite1} only applies to a Fock state. As an example, if one blindly applied that equation to a superposition of the degenerate space $|\psi\rangle=a|\uparrow\downarrow\uparrow\ldots\rangle + b|\downarrow\uparrow\downarrow\ldots\rangle $, the result,
$\mathcal{N}(\ell/2)=\left( |a|^2-|b|^2 \right) ^2/2\,,
$
is null for an equal superposition $a=b$. However, a direct  calculation of Eq.~\eqref{eq:apC(D)} reveals that the noise correlator matches the result $\mathcal{N}(\ell/2)=0.5$ associated to $|\uparrow\downarrow\uparrow\ldots\rangle$.

\item Going beyond on-site occupations, one can look into the case of a {bond-ordered state},
$\ev{\cdop_{r,\sigma}\cop_{s,\sigma'}}=\delta_{\sigma\sigma'}[\delta(s,r)n_{r\sigma}+0.5\delta(s,r+1) b_{r\sigma}+0.5 \delta(s,r-1)b_{r-1,\sigma}]$, where we index by $r$ the bonds connecting sites $(r,r+1)$. Eq.~\eqref{eq:cd_lattice} simplifies as,
$$
    \mathcal{N}(d)\approx \frac{1}{N^2+0.5B^2 \cos[Q(d)]}
    \big[-LN\delta(d)+\sum_{\sigma}\sum_{rs} e^{iQ(d)(r-s)} (n_{r\sigma}n_{s\sigma}+0.5b_{r\sigma}b_{s\sigma} )\big]\,,
    $$
where $B=\sum_{r\sigma}b_{r\sigma}\leq N$ denotes the total number of occupied bonds. For a distance of the form, $d=\nu \ell/2$, for $\nu \in \mathds{N}$, this reduces to,
    $$\mathcal{N}(\nu \ell/2)\approx
    \frac{\sum_\sigma \co{\sum_{r}(-1)^{\nu r}n_{r\sigma}}^2+0.5\sum_\sigma  \co{\sum_{r}(-1)^{\nu r}b_{r\sigma}}^2 }{(\sum_{\sigma r}n_{r\sigma})^2-0.5(\sum_{\sigma r}b_{r\sigma})^2}\,.$$

Considering now a state with a different occupation of even/odd bonds, $b_{e/o,\sigma}$, and sites, $n_{e/o,\sigma}$, one finds,
$$
    \mathcal{N}(\nu \ell/2)\approx
    \frac{\sum_\sigma \co{n_{e\sigma}+(-1)^\nu n_{o\sigma}}^2+0.5\sum_\sigma  \co{b_{e\sigma}+(-1)^\nu b_{o\sigma}}^2 }{\co{\sum_{\sigma}\pa{n_{e\sigma}+ n_{o\sigma}}}^2-0.5\co{\sum_{\sigma }b_{e\sigma}+ b_{o\sigma}}^2}\,,
$$
which reduces to Eq.~\eqref{eq:Fock_bipartite1} when no bonds are occupied.
\end{itemize}

\section{Tensor networks calculations}
\label{sec:TN}
The numerical sumations were performed using TenPy~\cite{hauschildEfficient2018}. The iMPS had maximum bond dimension of $\chi=200$ and the unit cell had length $L=2$. Ground states were found using iDMRG.  For faster convergence and symmetry breaking, no optimization was carried out in the first four sweeps, and we chose $|\uparrow \downarrow \uparrow \downarrow\rangle$ as the initial state for $V<2.4$, and $|\uparrow \downarrow - \uparrow\downarrow - \rangle$ for $V>2.4$. Particle number $N$ and spin $S_z$ were conserved.

\begin{figure}[!tbp]
    \centering   
    \includegraphics[width=0.6\columnwidth]{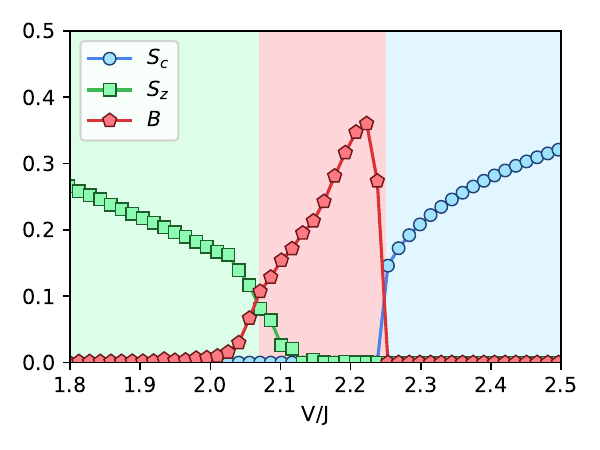}
    \caption{Zoom on the phase diagram of Fig.~\ref{fig:phases} for values of $V/J$ close to the BOW phase.}
    \label{fig:phasesZoom}
\end{figure}

Following this method, in Fig.~\ref{fig:phasesZoom} we have have calculated the LOPs associated to the phase diagram represented in Fig.~\ref{fig:phases}(e) for values of $V/J$ closer to the BOW phase. We observe that the LOP $B$, associated to this phase dominates in the range of parameters $V/J\in (2.07,2.25)$, which is the region we have coloured in red along this Letter.

\end{document}